\documentclass[aps,pra,showpacs,twocolumn,superscriptaddress,floatfix]{revtex4-1}

\usepackage{graphicx,color}
\usepackage{amsmath,amsthm,amsfonts,amssymb,bm}
\usepackage{times}
\usepackage{epsf}
\usepackage[colorlinks={true}]{hyperref}
\usepackage{bbm}

\usepackage[T1]{fontenc}
\usepackage[utf8]{inputenc}
\usepackage{ulem}
\usepackage{graphicx}
\usepackage{subfigure}

\hypersetup{citecolor={blue}, filecolor={blue}, linkcolor={blue}, urlcolor={blue}}

\begin{document}

\title{Thermal production, protection and heat exchange of quantum coherences}

\author{B. \c{C}akmak}
\email{bcakmak@ku.edu.tr}
\affiliation{Department of Physics, Ko\c{c} University, \.{I}stanbul, Sar\i yer 34450, Turkey}
\author{A. Manatuly}
\affiliation{Department of Physics, Ko\c{c} University, \.{I}stanbul, Sar\i yer 34450, Turkey}
\author{\"{O}. E. M\"{u}stecapl{\i}o\u{g}lu}
\affiliation{Department of Physics, Ko\c{c} University, \.{I}stanbul, Sar\i yer 34450, Turkey}

\begin{abstract}
We consider finite sized atomic systems with varying number of particles which have dipolar interactions among them and also under the collective driving and dissipative effect of thermal photon environment. Focusing on the simple case of two atoms, we investigate the impact of different parameters of the model on the coherence contained in the system. We observe that even though the system is initialized in a completely incoherent state, it evolves to a state with a finite amount of coherence and preserve that coherence in the long-time limit in the presence of thermal photons. We propose a novel scheme to utilize the created coherence in order to change the thermal state of a single two-level atom by repeatedly interacting it with a coherent atomic beam. Finally, we discuss the scaling of coherence as a function of the number of particles in our system up to $N=7$.
\end{abstract}


\date{\today}

\maketitle

\section{Introduction}

Vast majority of actual quantum systems are not isolated and in direct or indirect contact with their surrounding environments. The field of theory of open quantum systems deals with the problem of understanding the physics behind this mechanism and is a well established topic \cite{book1,book2,book3}. In general, interaction with the environment results in the loss of all quantum properties of the system, most important of all, coherence.

Many peculiarities of the quantum mechanics can be traced back to the wave-particle duality property of quantum particles. This dual behavior allows us to describe these particles as waves and put them in a coherent superposition of two (or more) possible states that they are allowed to occupy. Presence of such coherent superposition states actually is one of the the main differences between quantum mechanics and the classical mechanics. Despite its importance and many different manifestations, the genuine framework of characterization and quantification of coherence in an arbitrary quantum system is introduced only very recently \cite{baumgratz,levi,girolami,marvian}. The framework is formalized in \cite{baumgratz} by introducing a set of physically motivated conditions for a proper measure of coherence. Quantification of quantum coherence has attracted a lot of attention both on the fundamental level and about its applications in quantum critical, open and biological systems \cite{karpat,malvezzi,bromley,bhattacharya,damanet,guo,addis,zhang,behzadi,
review,vicente,streltsov,rana,yao,streltsov2015,brandner16,brandner17,uzdin,huelga,quantbio,scholes,chen}.

Quantum coherence is typically considered to be a resource for the quantum information devices \cite{review,vicente,streltsov,rana,yao,streltsov2015,brandner16,brandner17,uzdin}. More recently, it is understood that it can also be used as a “fuel” for quantum heat engines (QHEs) \cite{uzdin,scully,rosario,dag,turkpence,aberg,lostaglio,korzekwa}. Such a profound QHE, which can convert quantum coherence to useful work, can be practically appealing only if the abundant coherence is produced and protected either naturally or by energetically cheap artificial methods. Moreover it is necessary to be able to have a scheme that can harvest stored coherences as heat to produce work in a genuine heat engine cycle. Early studies, focusing on entanglement of a pair of two level atoms, suggest that a promising route towards natural and long-lived quantum coherence could be by simply using a thermal environment \cite{schneider,benatti,vedral,choi,ficek,brunner}. However such schemes yield too small quantum coherence. How to scale them up and how to convert them back to heat for realization of profound QHEs remain as open questions.

\begin{figure}[!t]
	\centering
	\begin{center}
		\subfigure[Heat is used to produce (``charge'') coherences]{
			\label{fig:fig2a}
			\includegraphics[width=7.25cm]{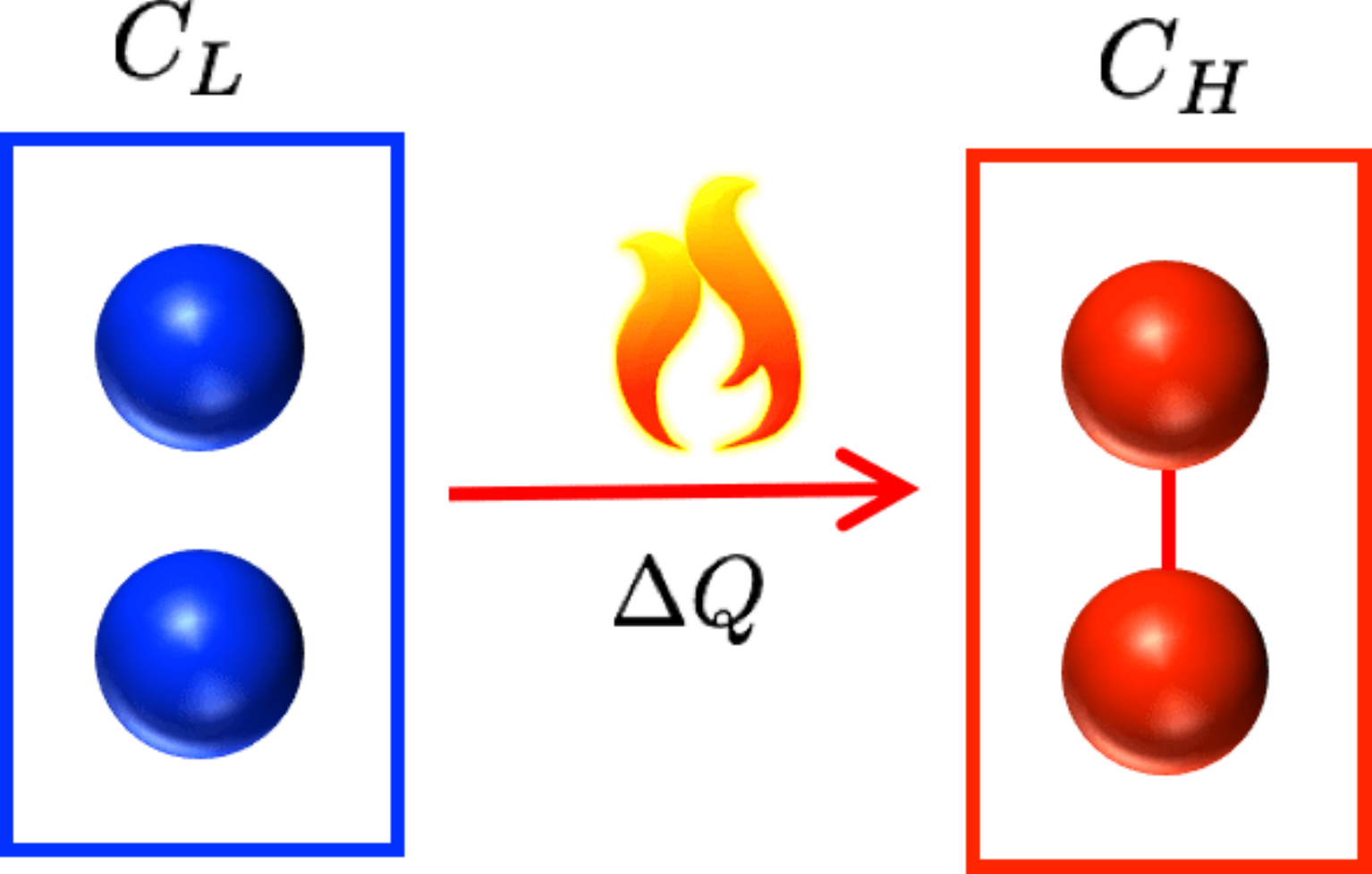}
		}\\
		\subfigure[Coherences are harvested back (``discharged'') as heat]{
			\label{fig:fig2b}
			\includegraphics[width=7.25cm]{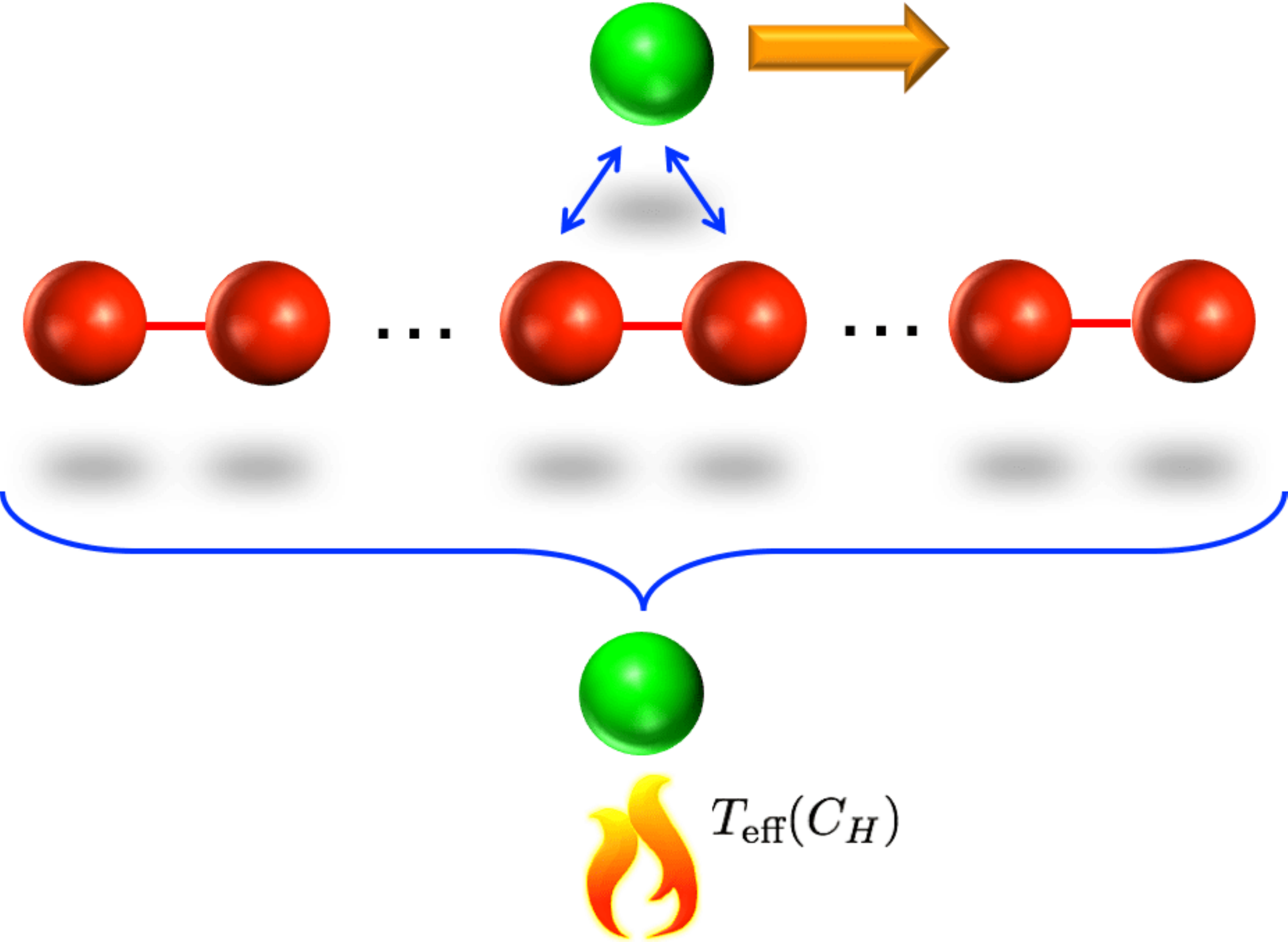}
		}\\
	\end{center}
	\caption{Schematic view of (a) creation and (b) harvesting of coherences in our model. A pair of two level atoms initially in a state with coherence 
	$C_L$ can be transformed into another state with higher coherence $C_H>C_L$ by collective interaction with a thermal bath. Energetic cost of
	generation of coherences is paid by the ``natural'' heat $\Delta Q$ withdrawn form the thermal bath. Some amount of this energy can be harvested
	back by a repeated interaction method where a single two level atom is coupled sequentially with a pair of atoms with coherence $C_H$. The 
	coupling happens at random times and the atom reaches a steady state eventually that can be described by a thermal state with an effective 
	temperature $T_{\text{eff}}$ that can be controlled by the $C_H$
	of the sub-environment atomic pairs.}
	\label{fig:fig2}
\end{figure}

In this paper, we propose that an ensemble of two level atoms can produce large amount of many body quantum coherence by collective coupling to a thermal environment. We find that the coherence exhibits a superlinear scaling with the number of ensemble atoms. In addition, we propose a scheme to harvest such thermally generated quantum coherence back as heat. We find that a single two level atom can be used as the working medium to harvest the coherences by randomly and repeatedly interacting it with similarly prepared coherent atomic clusters (pairs). The working atom reaches to a steady state that can be described by a thermal equilibrium state whose temperature depends on the coherence. This is in fact a generalization of a well-known route to thermalization by collision models \cite{scarani,ziman}, as well as the photo-Carnot engine in which the working fluid is the micromaser cavity field \cite{scully}. The intriguing point is that only certain coherences can be produced by collective heating and only those that can be converted back to heat. These coherences share the characteristic property of belonging to the energy degenerate subspaces (Dicke type states or Wigner-j matrix blocks in the computational basis) which is recently classified as heat exchange coherences \cite{dag}. Collective heating can "charge" such "flammable" coherences even if they are not present initially and preserve them in steady state so that they can be "discharged" back to heat by the initiation of harvesting scheme (cf.~Fig.~\ref{fig:fig2}).

The organization of the paper is as follows. We first describe our model system in Sec. \ref{model}. We then focus on the case of a pair of two level atoms and present key results on the quantum coherence generation and protection in thermal environment in Sec. \ref{results}. The subsections Sec. \ref{nbar} and Sec. \ref{dipoleint} will focus on the effects of the environment temperature and the dipolar coupling between the atoms, respectively. Sec. \ref{harvest} introduces our proposed scheme to harvest the thermally produced and stored coherences back as heat using a single two level atom. We generalize our results to the case of in multiple atoms in Sec. \ref{multiatom}. We conclude in Sec. \ref{conclusion}.

\section{Model}
\label{model}

The internal Hamiltonian of the atoms are given by $H_s=(\hbar \omega_0/2)\sum_{i=1}^N\sigma_i^z$, where $\sigma_i^z=|e_i\rangle\langle e_i|-|g_i\rangle\langle g_i|$ is the Pauli z matrix for the $i^{th}$ atom and $\omega_0$ is the transition frequency of the atom. For simplicity, we assume that all of the atoms in our sample act like
point-dipoles and polarized such that they all have the same dipole moment $\mathbf{d_{eg}^i}=\mathbf{d_{eg}}=\langle e|\mathbf{d}|g\rangle$. In the interaction picture associated with $H_s$, the master equation governing the dynamics of the atomic system is given by
\begin{equation}\label{me}
\frac{d\rho}{dt} = -\frac{i}{\hbar}[H_d, \rho]+\mathcal{D}_-(\rho)+\mathcal{D}_+(\rho)= \mathcal{L}(\rho),
\end{equation}
where the first term accounts for the unitary dipole-dipole interaction. Second and third terms are describing the spontaneous and thermally induced emission (dissipation), and thermally induced absorption (driving) processes respectively, whose explicit forms are as follows
\begin{equation}\label{dissipation}
  \mathcal{D}_-(\rho)=\sum\limits_{i,j=1}^N\gamma_{ij}(\bar{n}+1)(\sigma_j^-\rho\sigma_i^+-\frac{1}{2}\{\sigma_i^+\sigma_j^-, \rho\}),
\end{equation}
and
\begin{equation}\label{drive}
  \mathcal{D}_+(\rho)=\sum\limits_{i,j=1}^N\gamma_{ij}\bar{n}(\sigma_j^+\rho\sigma_i^--\frac{1}{2}\{\sigma_i^-\sigma_j^+, \rho\}).
\end{equation}
In the equations above $\sigma_i^+=|e_i\rangle\langle g_i|$ and $\sigma_i^-=|g_i\rangle\langle e_i|$ are the raising and lowering operators for the $i$th atom, $\bar{n}=(\exp(\beta\hbar\omega_0)-1)^{-1}$ is the mean number of thermal photons at the transition frequency of the atom at an inverse temperature $\beta$. $H_d=\hbar\sum_{i\neq j}f_{ij}\sigma_i^+\sigma_j^-$ is the dipole-dipole coupling Hamiltonian between the atoms in the considered system. The dipolar interaction strength $f_{ij}$ and the dissipation and driving rates $\gamma_{ij}$ are given as \cite{agarwal,stephen,lehmberg}
\begin{multline}
f_{ij}=\frac{3\gamma_0}{4}\left[(1-3\cos^2\alpha_{ij})\left(\frac{\sin\xi_{ij}}{\xi_{ij}^2}+\frac{\cos\xi_{ij}}{\xi_{ij}^3}\right)\right. \\ \nonumber 
\left. -(1-\cos^2\alpha_{ij})\frac{\cos\xi_{ij}}{\xi_{ij}}\right]
\end{multline}
and
\begin{multline}
\gamma_{ij}=\frac{3\gamma_0}{2}\left[(1-3\cos^2\alpha_{ij})\left( \frac{\cos\xi_{ij}}{\xi_{ij}^2}-\frac{\sin\xi_{ij}}{\xi_{ij}^3} \right)\right. \\ \nonumber 
\left. +(1-\cos^2\alpha_{ij})\frac{\sin\xi_{ij}}{\xi_{ij}} \right].
\end{multline}
Here, $\gamma_0=(\omega_0^3d_{eg}^2)/(3\pi\hbar\epsilon_0c^3)$ is the single atom spontaneous emission rate, $\xi_{ij}=k_0r_{ij}$ is a dimensionless parameter characterizing the distance between the particles with $k_0=\omega_0/c$ and $r_{ij}=|\mathbf{r_{ij}}|=|\mathbf{r_i}-\mathbf{r_j}|$ is the relative positions of the $i^{\text{th}}$ and $j^{\text{th}}$ atom. Finally, $\alpha_{ij}$ is the angle between $\mathbf{r_{ij}}$ and $\mathbf{d_{eg}}$.

The model admits two different regimes depending on the spatial distance of the atoms inside the ensemble. On one hand, we have the $\xi_{ij}\gg 1$ limit describing every atom is significantly distant with each other which results in the approximate model parameters, $f\approx 0$ and $\gamma_{ij}\approx \gamma_0\delta_{ij}$ for all $i$, $j$. In this regime we have no collective effects, every particle behave as independent. On the other hand, in the complete opposite limit of $\xi_{ij}\ll 1$ where $f_{ij}\approx (3\gamma_0(1-3\cos^2\alpha_{ij}))/4\xi_{ij}^3$ and $\gamma_{ij}\approx \gamma_0$ for all $i$, $j$, we are in the regime that collective effects play an important role. 

\section{Dynamics of coherence}
\label{results}

To begin with, we introduce the coherence measure that we are going to utilize to determine the amount of coherence in our system. It is called the $l_1$ norm of coherence and is just given by the sum of the absolute values of the off-diagonal elements in the density matrix of a given quantum system \cite{baumgratz}
\begin{equation}\label{l1norm}
  C_{l_1}(t)=\sum_{i\neq j}|\rho_{i,j}(t)|.
\end{equation}
$C_{l_1}$ is meaningful only if a reference basis for the density matrix is set and in what follows, we will fix our reference basis to be $\{ |e_1 e_2 \dots e_{n-1} e_n\rangle , |e_1 e_2 \dots e_{n-1} g_n\rangle , |e_1 e_2 \dots g_{n-1} e_n\rangle , \dots, \newline |g_1 g_2 \dots g_{n-1} g_n\rangle \}$, namely excitation/computational basis.

We may now proceed to analyze the time evolution of the $l_1$ norm under the dynamics dictated by the master equation of our physical model, Eq. (\ref{me}), for two atoms. We specifically concentrate the case of $N=2$, since it is the simplest ground in order to analyze the impact of different parameters in the model, on the time evolution of the atoms.

As previously mentioned, the dynamical model we consider here have two different regimes depending on the separation between the atoms. In the limit of $\xi_{ij}\gg 1$ for which the atoms are far apart from each other, we do not observe any interesting phenomena in the course of dynamics. Initial states that have no coherence do not accumulate any and the ones that have coherence lose it monotonically in finite time. In other words, individual coupling of atoms to the environment, generate no coherence and destroy the initially present amount. Therefore, we assume that the atoms in our system are spatially very close, corresponding to the $\xi_{ij}\ll 1$ limit where the collective effects are pronounced. Throughout this manuscript we will only consider identical dipole-dipole interactions between the atoms, $f_{ij}=f_0$. Furthermore, we make an implicit assumption that the mean number of photons, $\bar{n}$, appearing in Eq. (\ref{me}) is actually the mean number of photons that are at the transition frequency of the atoms, $\omega_0$, at a given temperature.

Before moving on to the multi-atom cases, we first consider the dynamics of coherence for a cluster constituted by two atoms in order to demonstrate interplay between the two main parameters in the system: the mean number of photons in the environment $n$ and the dipole-dipole interaction constant $f_0$. The conclusions we drew about the effect of these parameters also applies to larger atomic clusters. 

\subsection{Impact of mean number of thermal photons}
\label{nbar}

The discussion on the effect of mean number of photons in the environment surrounding the atoms is a bit more complicated to treat, even partially, for the most general initial state of two-atoms. Therefore, we need to choose an initial state and we have decided that initiating the dynamics from the ground state would generate the most suitable scenario to demonstrate the effect of thermal photons on coherence. If we take the initial state of the ensemble as the ground state of both atoms, $\rho(t=0)=|g_1g_2\rangle\langle g_1g_2|$ and evolve it according to the Eq.~(\ref{me}), we can calculate $\rho(t)$. We find that it is always in the form of a symmetric $X$-type (Dicke type) state with only two non-zero coherences lying in the central block. The coherences are always positive. $C_{l_1}$ as defined in Eq. (\ref{l1norm}) can be evaluated analytically. The result is independent of the dipolar interactions and given by
\begin{multline}\label{cohanalytic}
C_{l_1}(t)=\frac{\bar{n}(\bar{n}+1)}{3\bar{n}(\bar{n}+1)+1} \\  
-\frac{\bar{n}e^{-at}[(\bar{n}+1)\sqrt{\bar{n}(\bar{n}+1)}\cosh(bt)-\bar{n}^2\sinh(bt)]}{[3\bar{n}(\bar{n}+1)+1]\sqrt{\bar{n}(\bar{n}+1)}},
\end{multline}
where $a=2\gamma_0(2\bar{n}+1)$ and $b=2\gamma_0\sqrt{\bar{n}(\bar{n}+1)}$. We now analyze the results of different limits admitted by the above equation. 
\begin{figure}[h]
  \includegraphics[width=.44\textwidth]{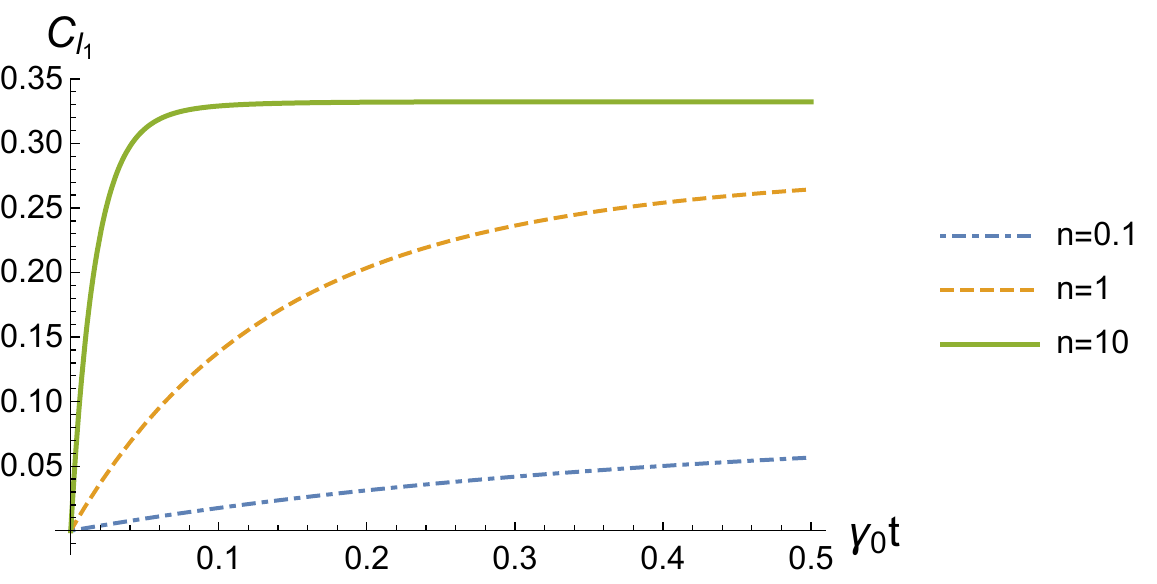}
  \caption{$C_{l_1}$ as a function of scaled time $\gamma_0t$ for different mean number of photons $\bar{n}$ in the environment for the two-atom initially in ground state.} \label{plotanalytic}
\end{figure}

To begin with, when there is no thermally induced processes, i.e. $\bar{n}=0$, $C_{l_1}$ remains zero independent of time. Since, in the absence of thermal environmental photons only spontaneous decay mechanism is present, an atomic system initiated in its ground state does not change during such dynamics. However, as soon as we have a finite temperature environment embodied by thermal photons, $\bar{n}\neq 0$, we begin to have non-zero coherence in the state of the system. As $t\rightarrow\infty$, the second term on the right-hand side of Eq. (\ref{cohanalytic}) goes to zero, leaving the first term unaffected which implies a time-invariant coherence in the system. In addition, in the case of high temperature environments where the mean number of photons is high, $n\gg 1$, $C_{l_1}$ converges to the value $1/3$. 

The mechanism behind the creation of coherence in the atomic ensemble is the thermal drive term in our physical model and this term is present only when we have $n\neq0$. The very presence of thermal photons in the environment surrounding our system, makes it possible for an atom to absorb that photon and make a transition to the excited state. Since we assume that the atoms in the ensemble are very close that they can be treated as they are indistinguishable. Therefore, it is not possible to identify the atom that absorbed the photon and ended up in the excited state which puts our system in a superposition state in the single (or more) excitation Hilbert space. However, the dissipation in the system, partially caused also by the same environment of thermal photons, opens up a channel that causes decays to the ground state, leading to loss of some of this created excitations. Nevertheless, in the long time limit, we see that our atomic ensemble comes to an equilibrium state, due to the trade-off between the loss and drive mechanisms, with a finite amount of coherence.

\subsection{Impact of dipolar interactions}
\label{dipoleint}

In order to investigate the impact of dipole-dipole interactions, we first determine the initial states whose evolution is influenced by $H_d$. For that aim, we have assumed that the system is disconnected from the outside environment, i.e. $\gamma_{ij}=0$, and it only evolves unitarily as determined by the first term in Eq. (\ref{me}). The evolution of elements of the denstiy matrix are presented in Eq. (\ref{app}), in the Appendix \ref{a1}. By inspection one can conclude that there are some initial states which are indifferent to the presence of dipolar interactions during the time evolution. These initial density matrices either have $\rho_{11}$, $\rho_{14}$ (and naturally $\rho_{41}$) and/or $\rho_{44}$ as their only non-zero elements, or satisfying the condition $\rho_{22}=\rho_{33}$ together with $\rho_{23}\in\Re$. Initial states lying outside the mentioned cases is affected by dipolar interactions and the effect is reflected to the dynamics of coherence as sinusoidal oscillations. Since the system evolution is unitary under $H_d$, such a behavior is natural. It is important to note that, when we consider the interactions with the environment, the Eq. (\ref{app}) describing the evolution of the density matrix elements will surely change and the effect of coherent dipole interactions will be suppressed due to open system dynamics. However, the dipolar interactions will affect the same density matrix elements apart from the cases outlined above, thus our conclusions above still holds true in the presence of interaction with the environment. 
\begin{figure}[h]
  \centering
  \includegraphics[width=.48\textwidth]{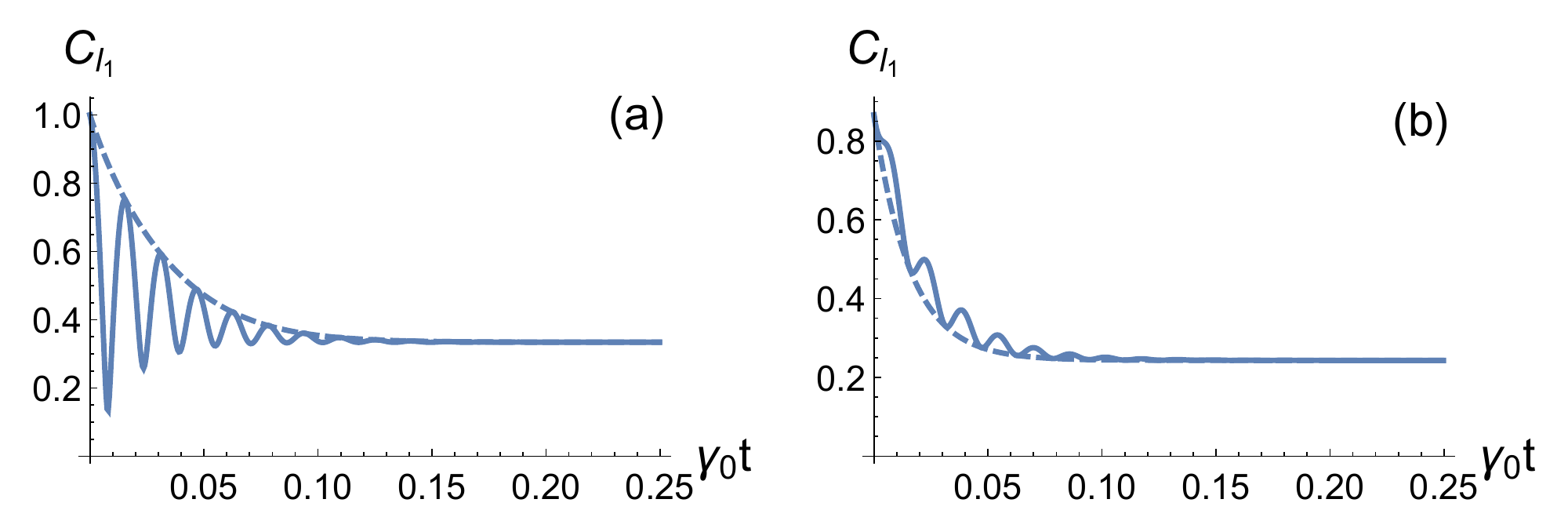}
  \caption{Dynamics of $C_{l_1}$ for the initial state (a) $|\psi(0)\rangle=(|ge\rangle+i|eg\rangle)/\sqrt{2}$ and (b) $|\psi(0)\rangle=(\sqrt{3}|ge\rangle+|eg\rangle)/2$. Solid and dashed lines represent $f_0/\gamma_0=10^2$ and $f_0/\gamma_0=1$ cases, respectively, with $\bar{n}=10$. Time is dimensionless and scaled with $\gamma_0$} \label{dipole}
\end{figure}

Fig. \ref{dipole} exemplifies the effect of dipole-dipole interactions for two different initial states for which the interaction does have an impact. It is possible to conclude that, in the presence of dissipation and drive, dipole interactions only have appreciable effects on the dynamics of coherence in short times: the value of coherence in the long-time limit does not depend on it. Moreover, to be able to see the effects in short times, the system must be in the regime $f_0/\gamma_0>1$ with $\bar{n}=10$, which are two competing energy scales in the dynamics. Even having $f_0/\gamma_0\approx 1$ is not enough to see an observable difference in the behavior of coherence.

\subsection{Harvesting the coherences}
\label{harvest}

In the previous sections we have seen that collective coupling of a pair of two level atoms to a heat bath allows for generation of certain coherences in the two atom density matrix. In this section we explore a scheme to harvest these coherences back as heat. For that aim, we consider a beam of atomic pairs carrying such coherences and assume that they interact with a two level atom at random time intervals. The total Hamiltonian describing the system can be written as follows

\begin{equation} 
 H=H_\text{q}+H_\text{b}+H_{\text{int}},
\end{equation}
where the single atom, an atom pair in the beam, and the interaction Hamiltonians are respectively given by
\begin{equation}
H_\text{q}=\frac{\hbar \omega _0}{2}   \sigma _0^z,
\end{equation}
\begin{equation}
H_\text{b}=\frac{\hbar \omega _0}{2} \sum\limits_{i=1}^2\sigma _i^z,
\end{equation}
\begin{equation}
H_{\text{int}}=\hbar g \sum\limits_{i=1}^2 (\sigma _i^+ \sigma _0^- + \sigma _i^- \sigma _0^+).
\end{equation}
Here $\omega _0$ is the transition frequency of the atoms, which are taken to be identical. The interaction coefficient is denoted by $g$. 

The atomic pairs arrive randomly at a rate $p$, and the interaction time $\tau$ is assumed to be short such that the condition $g\tau<<1$ is satisfied. In the interaction picture, the time-evolution operator is given by $U(\tau)=\exp(-i H_{\text{int}} \tau)$ whose exact expression, upto the second order in $(g\tau)^2$, is given in the Appendix \ref{a2}. The total density matrix of the system before each interaction is the product of the constituent density matrices $\rho (t)=\rho _{b} (t) \otimes \rho _q (t)$. Then, the master equation of the qubit can be written as
\begin{equation}
\dot{\rho}_q (t)=p\big[\sum\limits_{i,j=1}^N a_{ij} \sum\limits_{n=1}^N U_{ni} (\tau) \rho _q (t) [U_{nj} (\tau)]^{\dagger} -\rho_q (t)\big],
\end{equation}
where $a_{ij}$ are the density matrix elements of an atomic pairs in the beam $\rho _b (t)$. $U_{ni(j)}$ are the matrix elements of the time-evolution operator $U(\tau)$. It is important to note that, these matrix elements are actually operators in the single qubit Hilbert space. Expanding $U_{ni}$ and $U_{nj}$, we obtain the master equation in the Linbdlad form 
\begin{equation}
\dot{\rho}_q =-i [H_{eff},\rho _q]+\mathcal{L} _s \rho _q +\mathcal{L} \rho _q.
\end{equation}
The Hamiltonian $H_{\text{eff}}$ describes a coherent-drive term on the atom
\begin{equation}
H_{\text{eff}}=pg\tau (\lambda \sigma ^+ +\lambda ^* \sigma ^-).
\end{equation}
The Lindbladian $\mathcal{L} _s$ describes a squeezed reservoir effect on the atom and is expressed as
\begin{equation}
\begin{split}
\mathcal{L} _s \rho _q &=2\mu(\epsilon  \sigma ^+ \rho _q \sigma ^+ +\epsilon ^* \sigma ^- \rho _q \sigma ^-)\\
&=2\mu(\epsilon \mathcal{L} _s^e \rho _q +\epsilon ^* \mathcal{L} _s^d \rho _q),
\end{split}
\end{equation}
where $\mu=p(g\tau)^2$. Lindbladian $\mathcal{L} \rho _q$ is expressed by 
\begin{equation}
\mathcal{L} \rho _q=\mu(\frac{r_e}{2} \mathcal{L} _e  \rho _q +\frac{r_d}{2} \mathcal{L} _d  \rho _q),
\end{equation}
where excitation and de-excitation of the atom is described by
\begin{equation}
\mathcal{L}_e \rho _q=2\sigma ^+ \rho _q \sigma ^--\sigma ^- \sigma ^+  \rho _q- \rho _q \sigma ^- \sigma ^+,
\end{equation}
\begin{equation}
\mathcal{L}_d \rho _q=2\sigma ^- \rho _q \sigma ^+-\sigma ^+ \sigma ^-  \rho _q- \rho _q \sigma ^+ \sigma ^-.
\end{equation}
\\
\begin{center}
\begin{tabular}{|c|c|}
\hline
$r _e$ & $2a_{11}+a_{22}+a_{23}+a_{32}+a_{33}$ \\ 
$r _d$ & $2a_{44}+a_{22}+a_{23}+a_{32}+a_{33}$ \\
$\lambda$ & $a_{12}+a_{13}+a_{24}+a_{34}$ \\ 
$\epsilon$ & $a_{14}$\\ \hline
\end{tabular}
\end{center}
\begin{center}
\textbf{Table1.} The coefficients of the Lindbladians in the master equation
\end{center}
The Table 1 shows that coherences in the density-matrix of the atomic pair disjointly determine the contributions of the processes in the master equation. If the state of the atomic pairs is set so that $\lambda=0$, the master equation becomes
\begin{equation}
\dot{\rho}_q =\mathcal{L} _s \rho _q +\mathcal{L} \rho _q.
\end{equation}
By expressing $N=r _e/(r _d-r _e), \gamma=\mu (r _d-r _e)$, and $M e^{i \phi}=-2\epsilon \mu/\gamma $, we can rewrite the equation as follows
\begin{equation}
\begin{split}
\dot{\rho}_q=&\frac{1}{2}\gamma (N+1)(2\sigma ^- \rho _q \sigma ^+-\sigma ^+ \sigma ^-  \rho _q- \rho _q \sigma ^+ \sigma ^-)+\\
&+\frac{1}{2}\gamma N(2\sigma ^+ \rho _q \sigma ^--\sigma ^- \sigma ^+  \rho _q- \rho _q \sigma ^- \sigma ^+)-\\
&-\gamma M e^{i \phi}\sigma ^+ \rho _q \sigma ^+ -\gamma M e^{-i \phi}\sigma ^- \rho _q \sigma ^-.
\end{split}
\end{equation}
The master equation is the same as that of the two-level atom subjected to the squeezed thermal bath, with the Bloch equations \cite{gar} 
\begin{equation}
\label{bloch}
\begin{split}
\dot{\langle\sigma_x\rangle}= & -\frac{\gamma}{2}(2N+M+M^*+1)\langle\sigma_x\rangle\\
& -\frac{i\gamma}{2} (M-M^*)\langle\sigma_y\rangle+\frac{1}{2} ipg\tau(\lambda-\lambda^*)\langle\sigma_z\rangle,\\
\dot{\langle\sigma_y\rangle}= & -\frac{\gamma}{2}(2N+M+M^*+1)\langle\sigma_y\rangle\\
& -\frac{i\gamma}{2} (M-M^*)\langle\sigma_x\rangle-\frac{1}{2} pg\tau(\lambda+\lambda^*)\langle\sigma_z\rangle,\\
\dot{\langle\sigma_z\rangle}= & -\gamma[(2N+1)\langle\sigma_z\rangle+1]\\
& -2ipg\tau[(\lambda-\lambda^*)\langle\sigma_x\rangle+i(\lambda+\lambda^*)\langle\sigma_y\rangle],
\end{split}
\end{equation}
where $\gamma$ corresponds to vacuum spontaneous emission rate.

The nonunitary part of the master equation in fact could be transformed into a sum of two $\mathcal{L} _1$ and $\mathcal{L}_2$ dissipators in Lindblad form. We can transform them into two dissipators using $R_1$ and $R_2$ \cite{ban}
\begin{equation}
\begin{split}
\dot{\rho}_q&=-i [H_{\text{tot}},\rho _q]+\mathcal{L} _1+\mathcal{L}_2=\\&=-i [H_{\text{tot}},\rho _q]+\sum\limits_{i=1}^2(2R_i\rho_qR_i^{\dagger}-R_i^{\dagger}R_i\rho_q-\rho_qR_i^{\dagger}R_i)
\end{split}
\end{equation}
where 
\begin{equation}
\begin{split}
&R_1=\sqrt{\frac{\gamma(N_{\text{th}}+1)}{2}}R,\\
&R_2=\sqrt{\frac{\gamma N_{\text{th}}}{2}}R^{\dagger}, \\ 
&R=\sigma^-\cosh(r)+e^{i\phi}\sigma^+\sinh(r).\\
\end{split}
\end{equation}

The master equation is in the Lindblad form which is going to be of great use in the calculation of the heat current and work flux (power) in and out of our single qubit system due to the interaction with the atomic beam. As given in \cite{kos} the heat current and power are defined using the coherent evolution Hamiltonian and the dissipators as
\begin{equation}
J_i=\langle\mathcal{L}^* _i(H_{\text{tot}})\rangle=\text{Tr}\big[\rho_q\mathcal{L}^* _i(H_{\text{tot}})\big]
\end{equation}
\begin{equation}
P=\langle\frac{\partial H_{\text{tot}}}{\partial t}\rangle=\text{Tr}\big[\rho_q\frac{\partial H_{\text{tot}}}{\partial t}\big].
\end{equation}
Here they are defined in the laboratory frame, however we will calculate them in the interaction picture for convenience. Using $U=e^{i\omega_q\sigma_zt/2}$
\begin{equation}
\begin{split}
J_i&=\text{Tr}\big[\rho_q\mathcal{L}^* _i(H_{\text{tot}})\big]=\text{Tr}\big[U\rho_q\mathcal{L}^* _i(H_{\text{tot}})U^{\dagger}\big]=\\
&=\text{Tr}\big[U\rho_qU^{\dagger}U\mathcal{L}^* _i(H_{\text{tot}})U^{\dagger}\big]=\text{Tr}\big[\tilde{\rho_q}\tilde{\mathcal{L}}^* _i(UH_{\text{tot}}U^{\dagger})\big].
\end{split}
\end{equation}
where tilde denotes interaction picture term. Since the Hamiltonian in the interaction picture Eq.7 is given as 
\begin{equation}
H_{eff}=UH_{\text{tot}}U^{\dagger}-iU\frac{\partial U^{\dagger}}{\partial t}.
\end{equation}
The heat current is obtained as 
\begin{equation}
J_i=\text{Tr}\big[\tilde{\rho_q}\tilde{\mathcal{L}}^* _i(H_{eff}+\frac{1}{2}\omega_q\sigma_z)\big].
\end{equation}
Dropping tilde for convenience and denoting $\langle\sigma_z\rangle=\rho_{ee}-\rho_{gg}$, $\langle\sigma_+\rangle=\rho_{ge}$ 
 we calculate the heat current by directly using the definition of Eq.(17) as
\begin{equation}
\begin{split}
J_q & = J_1+J_2 \\
& =\frac{\gamma}{4}\omega_q(1-(2N+1)\langle\sigma_z\rangle)\\
 &-\frac{\gamma}{2}pg\tau\big[\lambda^*(\frac{2N+1}{2}\langle\sigma_-\rangle+\langle\sigma_+\rangle M)\\
 & +\lambda (\langle\sigma_-\rangle M^*+\frac{2N+1}{2}\langle\sigma_+\rangle)\big].
\end{split}
\end{equation}
\\
We can see from the above equation that $M$ can contribute heat flow only when $\lambda\neq 0$, which leads to non-thermal working qubit state. In other words, if $\lambda=0$, $M$ cannot influence the working qubit populations. Accordingly, $M$ consists o ineffective coherence that cannot be regarded as a heat exchange coherences. Only those coherences in $N$ can contribute to heat flow and temperature of the work qubit under the condition that the qubit is described by the canonical thermal state in equilibrium.

Let us now calculate the power. Using Eq.(14) in the interaction picture, similar to heat current, we calculate the power to be 
\begin{equation}
P=\frac{1}{2}\hbar \dot{\omega}_q\langle\sigma_z\rangle+pg\tau(\dot{\lambda}\langle\sigma_+\rangle+\dot{\lambda}^*\langle\sigma_-\rangle).
\end{equation}
In order to find the explicit forms of the heat current and power, we need to solve the Bloch equations. Their steady state solutions are presented in the Appendix(\ref{a3}). However, one can reach quick conclusions looking at the above expressions for heat current and power. The most important one is that when there is no coherent drive in the system there is no power received by the system qubit from the external source. Heat exchange coherences, the ones that is generated by the collective interaction of the atomic pairs with a heat bath, only contribute to the heat flux into the qubit system through the interactions with the atomic beam. As we will see next, this heat flux eventually leads to a change in the thermal state of the single qubit. 

The coherences that can be generated by the collective coupling of the atomic pair to the heat reservoirs lead to $\lambda=0$ and $\epsilon=0$. Specifically, only $a_{23}$ and $a_{32}$ can be generated (assuming initially the pair has no other coherence or no coherence at all). The master equation then becomes
\begin{equation}
\dot{\rho}_q=\mu(\frac{r_e}{2} \mathcal{L} _e  \rho _q +\frac{r_d}{2} \mathcal{L} _d  \rho _q).
\end{equation}
This equation can be solved exactly and the steady-state solution of the density matrix is given by
\begin{equation}
\rho_{ss}=
\begin{pmatrix}
\frac{r_e}{r_d+r_e} & 0\\
0 & \frac{r_d}{r_d+r_e}
\end{pmatrix}.
\end{equation}
We can assign an effective temperature for the atom as $T=-(\hbar \omega_0/k_b) \ln (r_e/r_d)$. Depending on $r_e$ and $r_d$ or the populations and coherences $a_{23}$, $a_{32}$ of the atomic pairs, the effective temperature can be negative or positive. At negative temperature population inversion occurs. For $r_e<r_d$ a well defined temperature exists and depends on coherences of the atomic pairs. In such a case, we conclude that thermally produced coherences are harvested back again as heat. Combination of collective heating and collisional harvesting schemes allows for exchange between quantum coherences and heat energy.

\section{Long-time behavior for different sized ensembles}
\label{multiatom}

\begin{figure}[h]
  \centering
  \includegraphics[width=.48\textwidth]{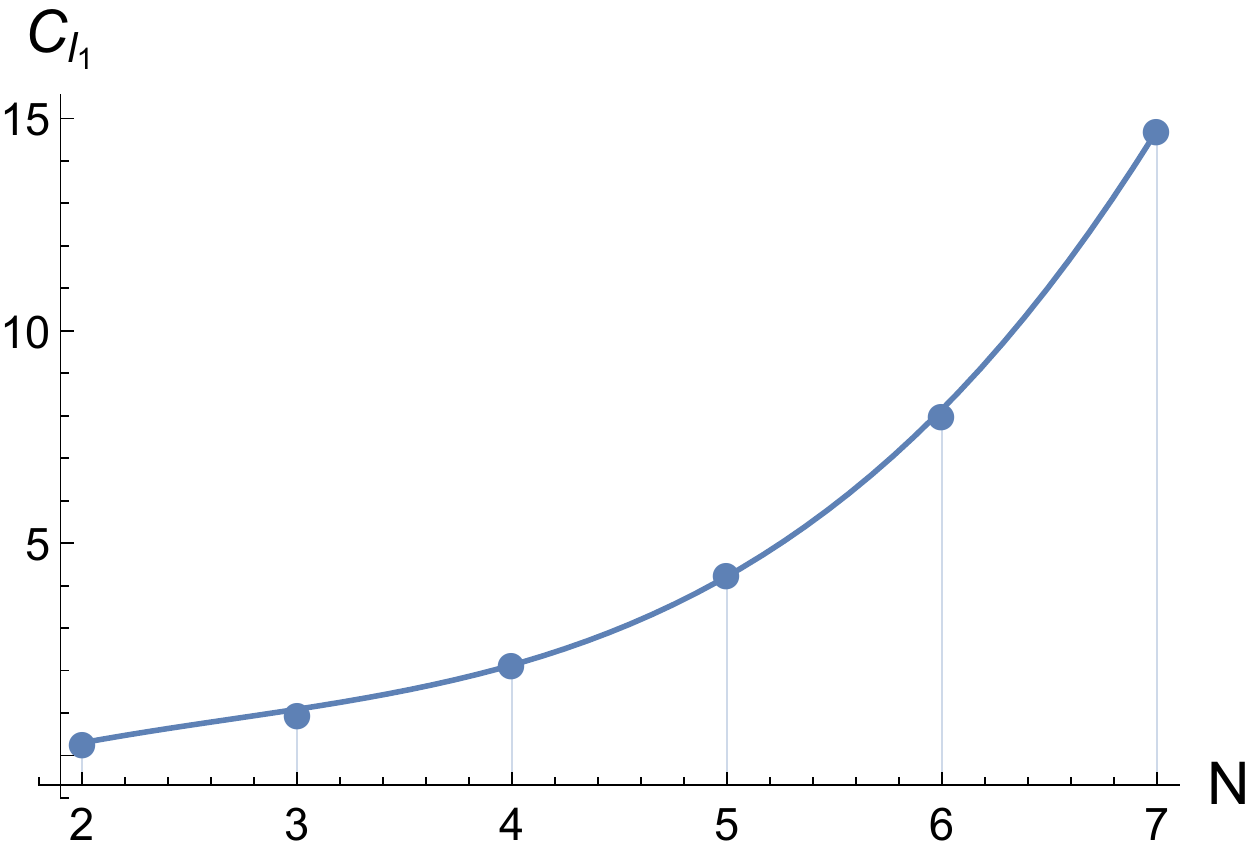}
  \caption{Scaling of $l_1$-norm of coherence in the long time limit with the number of atoms with an initial state where all the atoms are in their initial state. The dots represent the actual value of the coherence measure while the solid line is the cubic fit to these points. Model parameters are set to $f_0/\gamma_0=1$ and $n=10$. }\label{scaling}
\end{figure}

We consider the case of initiating all atoms in our ensemble in their ground state, i.e. $\rho(t=0)=|g_1 g_2 \dots g_N\rangle\langle |g_1 g_2 \dots g_N|$. Such a state has no coherence in our reference basis. However, looking at the dynamics of $C_{l_1}$ we observe that it increases monotonically with time and settles to a finite value in the long time limit, $C_{l_1}^{\text{lt}}$. The amount of coherence that is accumulated in the system increases with the increasing number of atoms in the ensemble, to be specific it shows a cubic best fit behavior, as presented in Fig. \ref{scaling}. The values of $f_0$ and $\gamma_0$ have negligible effect on the value that the coherence settles, however they can affect the time it takes to reach this particular value. $C_{l_1}^{\text{lt}}$ is only controlled by $\bar{n}$. It grows from zero with increasing $\bar{n}$ upto a certain value and than saturates.

The off-diagonal density matrix elements that contribute to the non-zero value of $C_{l_1}$ are the ones in the block diagonals adjacent to the main diagonal. To be more specific, if we group our reference basis according to the number of excitations, only matrix elements that are inside the symmetric subspaces of these groups have a finite value. These blocks are also known as the Jordan-Wigner blocks of a density matrix with a given basis. Smallest system of a pair of atoms have a state in the form of a symmetric $X$-type state structure with
real positive coherences. 
Larger ensembles still possesses only real positive coherences in the blocks along the main diagonal. In \cite{dag}, it was shown that the $\rho_{ij}$'s inside these blocks have a caloric value and are called heat-exchange coherences. This implies that, for example, when the atomic cluster is injected in a cavity, these coherences will change the temperature of the cavity and thermalize it to a different finite value. Their contribution to the thermalization
temperature is given by the addition of the coherences in Dicke type blocks. For real positive coherences the coherence measure $C_{l_1}$ is then
directly characterizes the caloric value of harvesting such coherences in micromaser type photonic quantum heat engines.

The initial state of the system that we consider in this section has no energy cost in the state preparation stage since all atoms are initiated in their ground state. The energy cost of generating coherences is paid by the thermal drive. If this stage can be done by utilizing natural thermal resources that can lead to reduced operational cost and hence increased efficiency of quantum thermal machine. 

\section{Conclusion}
\label{conclusion}

We have investigated the dynamics and the long-time behavior of coherence in an atomic cluster which have dipolar interactions between them, and also subject to dissipation and driving by the thermal photons in the environment. 
We began our discussion by trying to understand the effects of dipole-dipole interaction strength and the mean number of environmental photons on the dynamics of coherence. In order to present the explicit results analytically, we first consider a pair of atoms, then explore larger clusters, 
up to $7$ atoms, numerically. 

In the absence of thermal photons, $\bar{n}$, there is only dissipation on the atoms due to spontaneous emission processes whose rate is given by $\gamma_0$. Therefore, neither any coherence is generated nor the present coherence could survive in the joint state of the atoms on the course of the dynamics. In the case of a finite temperature environment, where $\bar{n}\neq 0$, system is driven by the photons, making it possible to generate and/or preserve the coherence. However, very presence of photons also induces thermal emission processes which creates a trade-off between dissipation and drive mechanisms. As a result the generated coherence cannot increase indefinitely and does not get affected by increasing $\bar{n}$ after a certain number.

The dipolar interactions have impact on some of the initial states, not all. When the time evolution is started from one of these states, coherence showed an oscillators behavior in short-times and it settled to the value determined by the dissipation and drive terms. In other words, long-time value of coherence is indifferent to the dipole interactions. Moreover, the interaction strength ,$f_0$, must be greater than the $\bar{n}\gamma_0$ to be able to have this effect on the system in short-times. 

Furthermore, we have calculated the scaling behavior of coherence in the long-time limit for atomic clusters upto $N=7$. We chose to initiate all atoms in their ground state for which there is no coherence present in the system. However, during the dynamics certain amount of coherence builds up in the cluster, saturates and becomes time-invariant for the rest of the evolution. Creation of coherence in the ensemble due to the fact that atoms are considered to be very close to each other. An excitation created by absorption of a photon which is made possible by the drive mechanism, delocalizes throughout the system and puts it in a superposition state of all single excitation subspace. Keeping in mind the losses in the system, coherence cannot grow to its maximum, but gain and loss balance each other to leave the system in a state with finite coherence.

We would like to emphasize that creation and preservation of coherence in the model considered here, has its explanation in the collective quantum effects taking place in our atomic cluster due to their proximity. Other explanations of enhanced coherence life-time such as non-Markovianity \cite{addis,zhang,behzadi,guo} or decoherence-free subspaces \cite{dfs} is not the case for our model. The former requires negative decay rates, $\gamma_{ij}<0$, to be present and the latter shows up in the pure dephasing dynamics of the subject system, both are features that our system do not possess.

Fundamentally we presented the heat equivalent of certain quantum coherences in atomic clusters by identifying means of mutual exchange between heat and quantum coherences. Accordingly motive power of quantum coherence can be defined for operation of quantum machines
between quantum reservoirs with a coherence gradient.
Our results can also be practically significant for reducing operational costs of quantum heat engines by 
utilizing natural quantum coherence resources and for designing multi qubit
quantum machines where both the resource and working fluid parts consists of qubits, reducing the interfacing challenges. Searching existence of
such mechanisms in biological systems or utilizing them for photovoltaic applications can be envisioned.

\begin{acknowledgments}
The authors acknowledge support from a University Research Agreement between Lockheed-Martin Corp. and Koç University. The authors would like to thank J. Anders, M. Esposito and M. Campisi for fruitful discussions.
\end{acknowledgments}

\onecolumngrid

\appendix

\section{Effect of dipolar interactions}
\label{a1}

When only coherent dipole-dipole interaction is present, i.e. in the absence of drive and dissipation which corresponds to $\gamma_0=0$ case, the elements of the density matrix of two atom ensemble evolves as follows
\begin{align}\label{app}
\rho_{11}(t) &= \rho_{11}(0) \\ \nonumber
\rho_{12}(t) &= \rho_{12}(0)\cos(ft)+i\rho_{13}(0)\sin(ft) \\ \nonumber
\rho_{13}(t) &= \rho_{13}(0)\cos(ft)+i\rho_{12}(0)\sin(ft) \\ \nonumber
\rho_{14}(t) &= \rho_{14}(0) \\ \nonumber
\rho_{22}(t) &= [(\rho_{22}(0)+\rho_{33}(0))+(\rho_{22}(0)-\rho_{33}(0))\cos(2ft)+2i\Im[\rho_{23}]\sin(2ft)]/2 \\ \nonumber
\rho_{23}(t) &= [\Re[\rho_{23}(0)]+i(2\Im[\rho_{23}(0)]\cos(2ft)+(\rho_{22}(0)-\rho_{33}(0))\sin(2ft))]/2 \\ \nonumber
\rho_{24}(t) &= \rho_{24}(0)\cos(ft)-i\rho_{34}(0)\sin(ft) \\ \nonumber
\rho_{33}(t) &= [(\rho_{22}(0)+\rho_{33}(0))-(\rho_{22}(0)-\rho_{33}(0))\cos(2ft)-2i\Im[\rho_{23}(0)]\sin(2ft)]/2 \\ \nonumber
\rho_{34}(t) &= \rho_{34}(0)\cos(ft)-i\rho_{24}(0)\sin(ft) \\ \nonumber
\rho_{44}(t) &= \rho_{44}(0), \\ \nonumber
\end{align}
where $\rho_{ij}(0)$'s are their initial values. By looking at the above equations, we can determine which initial states will not get affected by the dipolar interactions during the course of time. One can immediately see that states residing in the $\rho_{11}$, $\rho_{14}$ (and naturally $\rho_{41}$ which is $\rho_{14}^*$) and $\rho_{44}$ subspace initially, do not feel the presence of dipole-dipole interaction. Moreover, states having $\rho_{22}(0)=\rho_{33}(0)$ and $\rho_{23}(0)\in\Re$ will also be indifferent to the dipolar interactions.

\section{Time evolution operator}
\label{a2}

The time-evolution operator is given as follows
\begin{equation}
U(\tau)=\text{exp}(-i H_\text{int} \tau)=\text{exp}(-ig\tau \sum\limits_{i=1}^2 (\sigma _i^+ \sigma _0^- + \sigma _i^- \sigma _0^+)),
\end{equation}
where we took $\hbar=1$. We need to evaluate this operator upto the second order in $(g\tau)^2$. Thus if we set $S=\sum\limits_{i=1}^2 (\sigma _i^+ \sigma _0^- + \sigma _i^- \sigma _0^+)$, the evolution operator to the second order in $(g\tau)^2$
\begin{equation}
U(\tau)\approx \mathbbm{1}-ig\tau S-\frac{(g\tau)^2}{2}S^2.
\end{equation}
 The collective raising and lowering operators are given as 
\begin{equation}
\sum\limits_{i=1}^2 \sigma _i^+=
\begin{pmatrix}
0 & 1 & 1 & 0\\
0 & 0 & 0 & 1\\
0 & 0 & 0 & 1\\
0 & 0 & 0 & 0\\
\end{pmatrix}, 
\qquad  \sum\limits_{i=1}^2 \sigma _i^-=
\begin{pmatrix}
0 & 0 & 0 & 0\\
1 & 0 & 0 & 0\\
1 & 0 & 0 & 0\\
0 & 1 & 1 & 0 
\end{pmatrix},
\end{equation}
which give us the $S$ operator in the following form
\begin{equation}
S=
\begin{pmatrix}
0 & \sigma _0^- & \sigma _0^- & 0\\
\sigma _0^+ & 0 & 0 & \sigma _0^-\\
\sigma _0^+ & 0 & 0 & \sigma _0^-\\
0 & \sigma _0^+ & \sigma _0^+ & 0 
\end{pmatrix}.
\end{equation}
Finally the time-evolution operator is evaluated as
\begin{equation}
U(\tau)=
\begin{pmatrix}
1-(g\tau)^2\sigma _0^- \sigma _0^+ & -ig\tau\sigma _0^-  & -ig\tau\sigma _0^- & 0\\
-ig\tau\sigma _0^+ & 1-(g\tau)^2 /2 & -(g\tau)^2 /2 & -ig\tau\sigma _0^-\\
-ig\tau\sigma _0^+ & -(g\tau)^2 /2 & 1-(g\tau)^2 /2 & -ig\tau\sigma _0^-\\
0 & -ig\tau\sigma _0^+ & -ig\tau\sigma _0^+ & 1-(g\tau)^2\sigma _0^+ \sigma _0^-\\
\end{pmatrix}.
\end{equation}

\section{Steady-state solutions of the Bloch Equations}
\label{a3}

The steady-state solutions of Eq.s \ref{bloch} are found as follows
\begin{equation}
\hspace*{-4cm}
\begin{split}
\langle\sigma_x\rangle_{ss}= & \frac{\frac{i\gamma pg\tau}{4}[(2N+M+M^*+1)(\lambda-\lambda^*)+(M-M^*)(\lambda+\lambda^*)]}{d},\\
\\
\langle\sigma_y\rangle_{ss}= & \frac{\frac{-\gamma pg\tau}{4}[(2N+M+M^*+1)(\lambda+\lambda^*)-(M-M^*)(\lambda-\lambda^*)]}{d},\\
\\
\langle\sigma_z\rangle_{ss}= & \frac{\frac{\gamma^2}{4}[(2N+M+M^*+1)^2+(M-M^*)^2]}{d},\\
\end{split}
\end{equation}
where $d$ is given as 
\begin{equation}
\begin{split}
d=&\big[(pg\tau)^2((M-M^*)(\lambda^2-\lambda^{*2})-2|\lambda|^2(2N+M+M^*+1))\\
&-\frac{\gamma^2(2N+1)}{4}((2N+M+M^*+1)^2+(M-M^*)^2)\big].
\end{split}
\end{equation}



\end{document}